# Validating Drone Trust Testing in Navigation Deviation Cases in Simulation

Zahra Rezaei Khavas, Edwin Meriaux, Amin Majdi, Paul Robinette
Department of Electrical and Computer Engineering
University of Massachusetts Lowell
Lowell, Massachusetts
Email: Edwin_Meriaux@student.uml.edu

Edwin Meriaux and Zahra Rezaei Khavas are Co-first authors on this paper

*Abstract*—
**Developing videos for trust testing is very time-consuming, expensive, and potentially dangerous. For trust tests, it requires a person to be flying the drone while another might be filming. The drones can be very expensive and if something goes wrong the costs might be very high. In previous work, we have looked at how collisions and basic communication loss can be accurately modeled in simulation and to be able to generate the same trust results from users. That work looked at two specific cases using two drones, but to expand upon this in other cases more testing is required. This paper looks to propose how to test and evaluate the change in a user's trust of a drone when it is experiencing path deviation in simulation. If the environment is very realistic can simulations be a good alternative to real life videos for trust testing when there is path deviation. This deviation can occur due to the physical conditions of the space, faulty piloting, or communications loss.**

*Index Terms*—**UAVs, sUAS, drones, human-drone interaction, human-drone trust, Human-robot trust, drone failure, drone communication loss, packet loss, correlated packet loss, path deviation, drone collision, airsim, simulation,**

## I. INTRODUCTION

Evaluating a drone encompasses many different aspects of the robot's capabilities. Its navigation accuracy, obstacle avoidance, communication, and trust all need to be tested to make sure the drone meets the necessary parameters. In open environments, work has been conducted by National Institute of Standards and Technology (NIST) to develop tests that span multiple domains to have a full understanding of the capabilities of any one drone [1]. More work has been done evaluating the effects of collision between people and drones ([2]). In previous phases of our work at the University of Massachusetts Lowell (UML) we have done a lot of work on the individual aspects of drone evaluation for underground environments [3–7]. These tests can be very costly and time consuming to run. There are many cases where expensive drones can be damaged possibly causing the end of any more testing. This means that there is a need for testing drones and for testing drones as quickly, cheaply and accurately as possible. To make testing quick and cheap, simulations work, if done correctly, can be leveraged. This would allow for many tests to be run without damaging the drone and the speed they can be done at is just limited by the speed of the computer being used. If the simulation is done accurately the results should be very similar to in real life. In our previous work we have looked at the intersection of trust and communication loss [8]. This started with the in person tests where videos of the test were made while the drones were experiencing communication loss or collisions and then users were given questionnaires to evaluate their trust in those drones. This was then repeated in simulation to validate that simulation results could accurately mimic real life trust results in those cases. The goal of this paper is to work on the intersection between trust and path deviation. As a drone operates its required mission a large number of possible events can occur that cause deviation in the operation. If a user is doing work near this drone they must be confident that the drone will not harm them. If they lose trust they might not operate at their peak capabilities. Path deviation is a common occurrence for drones and can potentially harm people nearby. This means that understanding how much deviation will affect trust is critical to understanding when a drone can be used in a given space. This proposed work will look at how well the simulated results can mimic the real life ones to reduce the need for real life tests. In the previous simulation work we have used Airsim which is a Microsoft developed drone simulator [9]. This simulator has been used by other projects in evaluating human robot trust [10, 11]. The reason it is used is due to the very realistic visuals and physics in AirSim since it is built on the Unreal Engine.

## II. METHODOLOGY

This experiment is based on the comparison between features and capabilities of two platforms and the effects of those on human evaluation of drones trustworthiness in case of different drone failures. We based our experiment on assessing the effects of differences in the physical appearance and capabilities of two drones namely Elios and Mavic drone in this experiment. We used the same two drones in our previous experiment ([3, 4, 8]) and developed a strategy to assess the effects of different failures by these two drones on human trust. The proposed work will continue to use those platforms to maintain consistency. This phase will be centered on evaluating the human user's trust in drones when the drone deviates from its path. The goal is to validate as the deviation increase it can be properly modelled in simulation. This validation will happen by running the tests in the real world, filming the experiments, having people rate their trust of the different videos, build the simulation counterparts and then evaluating the trust results in each case. If the trust results are compatible, that would indicate that these cases can be simulated instead of running in the real world. This is useful because it allows for quicker testing and the ability for the drones to not be damaged during testing.

*A. Actual and Simulated Drones*

Two real and two simulated drones have been used in this experiment, in the following different characteristics of these drones are described in details.

- Elios Drone [12] (see Figure 1 and 2): Made by Flyability the Elios 2 is a 4 motor drone that is protected inside a cage. The noise it makes is high pitch and high tone.
- Mavic Drone [13] (see Figure 1 and 2): Made by DJI. This is a drone with 4 motors that has no propeller guard. The noise it makes is low pitch and low tone.
- The way the drones are simulated is twofold. The first is visual. Either a 3D model of the drone is procured from the manufacturer (or from any 3rd party that has modelled it) or, if this does not exist, the model is generated using a CAD program like Solidworks. Then this is brought in to the simulator, where the drone is modified to have the specific physical parameters that it needs to fly as accurately as possible. This includes the thrust metrics of the drone, which determine the acceleration and deceleration the drone will endure.

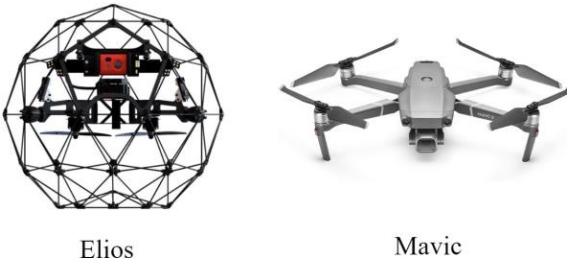

Fig. 1. The real drones platforms used in this experiment

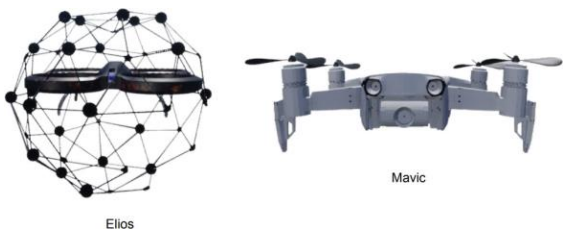

Fig. 2. The simulated drones platforms used in this experiment were made to replicate the platforms in Figure1

*B. Test Environment and Drones*

The test environment is built in Airsim[14]. This is a drone simulator developed by Microsoft on top of the Unreal Engine[15]. For the sake of previous tests in, a full environment has been built. It is a replica of the UMass Lowell's New England Robotics Validation and Experimentation (NERVE) Center[16] (this was the space used for the real world testing so far). The replica is identical visually and physically. It can be seen in Figure 3 and the drone's First Person View (FPV) can be seen in Figure 4. A comparison of the simulated space compared to the real world space can be seen by comparing 5 and 6.

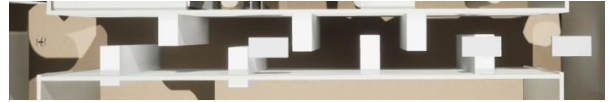

Fig. 3. Top Down View in the simulated NERVE space [3]

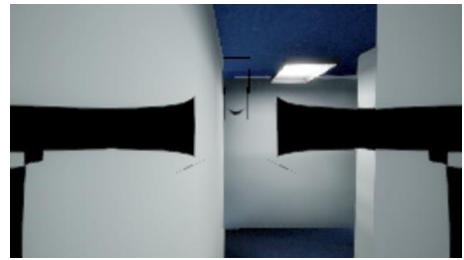

Fig. 4. The simulated drones platforms used in this experiment made to replicate the platforms in Figure1 [3]

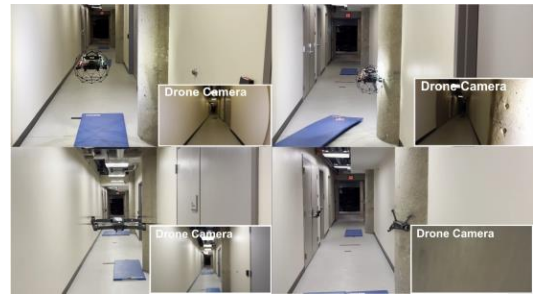

Fig. 5. Real world test videos. Elios's collision is on the top row and Mavic's collision is on the bottom row.

*C. Experiment Design*

There are three components to the experiments. The first is the path deviation, the second is the cause of the path deviation, and the third is the trust evaluation.

*1) Path Deviation and Collision:* As path deviation is increase for a drone, it will move further and further from its designated path. Four different cases of path deviation will be tested. They differ from each other in terms of severity of the deviation. The first one is no path deviation, as can be seen in Figure 7. The drone flies exactly the designated path. The second is minor path deviation (Figure 7), the third is major path deviation (Figure 8), and the last path deviation to the point of collision (Figure 8). Minor deviation will be defined as deviation less than 0.25 meters, while major deviation will be anything greater than that. The fourth case will be a case with collisions. This case is more difficult to simulate. How the drone falls to the ground must be properly modelled for the simulations to be believable. If they are not then the trust values in simulation will not properly show the trust values in real life, making the information effectively useless.

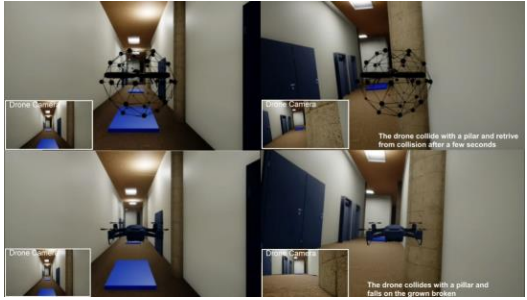

Fig. 6. Simulated test videos. Elios's collision is on the top row and Mavic's collision is on the bottom row.

*2) Deviation Causes:* There are many reasons why the path deviation in the previous section would occur in real life. They can be due to communication loss, due to the space the agent is flying in, or due to the pilot/algorithm flying the drone. These are just 3 possible causes of the deviation. All of these can be modelled in simulation. The physics of the space can allow for air currents, communication loss can be generated to cause path deviation, and faulty flight paths can be given. As the causes of the deviation increase in severity, then the deviation generated will increase and possibly lead to collisions.

*3) Trust Evaluation:* When the drones are tested for trust, videos of the flight are needed. The videos replace the need for the user to be in the physical space. Bringing potentially hundreds of evaluators into the space one at a time can be dangerous to the people, time-consuming, and very costly. By showing videos of the test instead, this process is facilitated. Once the videos are shown, the evaluators then answer some questions from questionnaires such as HTCM and JIAN to be used ([17] and [18]) to get an understanding of trust values. The user will first be told the situations the drones are in. They are flying through a tunnel to map it out, but there is path deviation and this can cause problems for the drones. The user will either be shown a pair of videos in simulation or in real world. Each video will contain a different drone (one will have Elios and the other Mavic). The participants will be shown both drones in the same state. The person will then be told that, at random, they will work with one of the drones to conduct the mapping operations in the space. The user must answer the trust questions based on the videos they saw to quantify how well the drone they were assigned. The order of the videos shown will also be randomized.

The users will be procured through Amazon's Mechanical Turk (MTurk) [1] or Prolific[2]. While the HTCM and JIAN questionnaires will be used, not all the questions will be. The two questionnaires' questions are not all necessarily good in this specific case. Questions one through three in Jian's set cannot be used because they do not relate to this experiment. The unfilled question in HTCM are filled with "the drone". For questions 7 through 9 the unfilled spots are replaced with "the mapping mission".

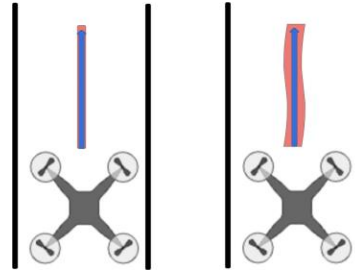

Fig. 7. Diagram on the left has the drone flying with no flight deviation and on the left the drone has small path deviation. The black line is the drone's intended path while the red line is the possible deviations the drones might have.

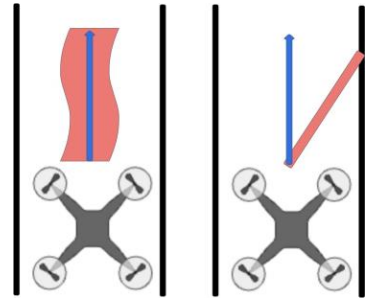

Fig. 8. Diagram on the left has the drone flying with major flight deviation and on the left the drone has collided into the wall. The black line is the drone's intended path while the red line is the possible deviations the drones might have.

### D. Videos

There will exist 16 videos total. They are divided by which drone is being flown, whether the drone is flying in real world or in simulation, and what is the severity of the deviation or if it is a collision. This is broken down in the following section.

1) Scenario 1: No path deviation. The first case has no packet loss in simulation or in real world. The flights should not have deviations or collision. This section contains 4 videos (2 for Elios and 2 for Mavic).
   a) Modality 1: The videos are from real world with no path deviation.
   b) Modality 2: The videos are from simulation with no path deviation.

[1]https://www.mturk.com/
[2]https://www.prolific.co/

2) Scenario 2: Minor path deviation. The Mavic and Elios drones are flown in the space with only minor deviation in the flight path. This section contains 4 videos (2 for Elios and 2 for Mavic).
   a) Modality 1: The videos are from real world with minor path deviation.
   b) Modality 2: The videos are from simulation with minor path deviation.
3) Scenario 3: Major path deviation. The Mavic and Elios drones are flown in the space with major deviation in the flight path. This section contains 4 videos (2 for Elios and 2 for Mavic).
   a) Modality 1: The videos are from real world with major path deviation.
   b) Modality 2: The videos are from simulation with major path deviation.
4) Scenario 4: Wall collision. The Mavic and Elios drones are flown in the space and repeatedly collide into the wall while deviating from the path. This section contains 4 videos (2 for Elios and 2 for Mavic).
   a) Modality 1: The videos are from real world with the wall collision.
   b) Modality 2: The videos are from simulation with the wall collision.

*E. Hypothesis*

All 16 videos are to be used in the goal of showing one hypothesis. In the case of a drone deviating from its path, if the drone, its flight and the space it is in is correctly modelled then the change in the trust a user will have for the drone can be properly evaluated in MTURK or Prolific using simulated videos of the drone instead of real life video tests. This means that not only must the space and the drone be physically and visually accurate, but the flight and collision must also be. If the drone crash is not properly implemented visually then it will not allow users to properly evaluate the trust of that drone. The trust values they give will not correctly correspond to the drone in question. A way this can be fixed in the simulator is by manipulating the underlying code in the simulator Airsim or by manipulating just the visuals in the video shown to the user. Each drone must fly and fall accurately in simulation. The trust results will be used to validate the hypothesis by calculating the associated p-values of trust results. If the p-value is high, then the alternative hypothesis that the trust values are very different can be rejected. If the tests are very different, that would mean the trust values are below $\alpha = 0.1$.

## III. CONCLUSION

Based on the prior work we have conducted in this realm, we expect that the results will show that the drones, when realistically simulated in a well modelled space, will have similar trust results in real world and in simulations when experiencing path deviation. This work will be a step in our larger project and goals to conduct simulated testing for drone evaluation. Every step requires validation testing to show that drone responses similar in our simulations compared to in real world. This work shows that high levels of realism are necessary and critical when social navigation is being conducted to make sure the humans trust the videos accurately.